\documentclass[twocolumn,showpacs,superscriptaddress,prl,aps]{revtex4}
\usepackage{graphicx}
\usepackage{epstopdf}
\usepackage{subfigure}
\usepackage{color}

\begin{document}
\title{BaMn$_2$P$_2$: Highest magnetic ordering temperature 122-pnictide compound}
\author{B. S. Jacobs}
\altaffiliation{sjacobs@uj.ac.za}
\affiliation {Department of Physics, University of Johannesburg, Johannesburg 2006, Gauteng, South Africa}
\author{Abhishek Pandey}
\altaffiliation{abhishek.pandey@wits.ac.za}
\affiliation {Materials Physics Research Institute, School of Physics, University of the Witwatersrand, Johannesburg 2000, Gauteng, South Africa}
\affiliation {Department of Physics and Astronomy, Texas A$\&$M University, College Station, Texas 77843, USA}

\date{\today}

\begin{abstract}

We report the growth of high-quality single crystals of ThCr$_2$Si$_2$-type tetragonal BaMn$_2$P$_2$ and investigation of its structural, electrical transport, thermal and magnetic properties. Our results of basal plane electrical resistivity and heat capacity measurements show that the compound has an insulating ground state with a small band gap. Anisotropic susceptibility $\chi_{ab,c}(T)$ data infer a collinear local-moment N\'eel-type antiferromagnetic (AFM) ground state below the ordering temperature $T_{\rm N} = 795(15)$~K, which is highest among all the ThCr$_2$Si$_2$- and CaAl$_2$Si$_2$-type 122-pnictide compounds reported so far suggesting that the strength of magnetic exchange interactions is strongest in this material. The magnetic transition temperatures of BaMn$_2$$Pn_{2}$ ($Pn$ = P, As, Sb, Bi) compounds exhibit a monotonic decrease with the increase of tetragonal unit cell parameters $a$ and $c$, suggesting a strong dependence of the strength of the decisive magnetic exchange interactions on the separation between the localized spins residing on the Mn-ions. The observed monotonic increase of both $\chi_{ab}$ and $\chi_{c}$ for  $T > T_{\rm N}$ suggests that short-range dynamic quasi-two dimensional AFM correlations persist above the $T_{\rm N}$ up to the highest temperature of the measurements. The large $T_{\rm N}$ of BaMn$_2$P$_2$ demands for systematic hole-doping studies on this material as similar investigations on related BaMn$_2$As$_{2}$ with $T_{\rm N} = 618$~K have led to the discovery of an outstanding ground state where AFM of localized Mn-spins and itinerant half-metallic ferromagnetism with $T_{\rm c} \approx 100$~K originating from the doped holes coexist together. 

\end{abstract}

\maketitle

\section{Introduction}

The search for new transition metals-based superconductors and their probable parent compounds got intensified after the discovery of high-temperature (high-$T_{\rm c}$) superconductivity (SC) in iron-based layered pnictides and chalcogenides \cite{Wang-2008,Paglione-2010,Johnston-2010,Stewart-2011}. Specifically, after the discovery of SC in doped-BaFe$_2$As$_2$, significant efforts were put into the investigations of ThCr$_2$Si$_2$-type 122-tetragonal systems that crystallize with space group $I4/mmm$. The investigations carried out so far suggest that iron is an essential ingredient for achieving high-$T_{\rm c}$ SC within this family of compounds. On the other hand, exploratory efforts on related systems where in place of iron some other 3$d$ or 4$d$ transition metal ions constitute the metal-pnictide sublattice have led to quite interesting outcomes and ground states such as prototypical half-metallic behavior \cite{Pandey-2013a,Ueland-2015,Pandey-2015} spin-liquid phase embedded with antiferromagnetic (AFM) and ferromagnetic (FM) fluctuations \cite{Pandey-2013b,Jayasekara-2013,Li-2019}, highly frustrated itinerant magnetism \cite{Sapkota-2017}, low-temperature SC \cite{Anand-2013,Anand-2014,Wang-2015, Zhao-2020}, discovery of new layered magnetic phases \cite{Pandey-2018}, large negative magnetoresistance (MR) and Anderson localization \cite{Huynh-2019,Ogasawara-2021,Ogasawara-2022} to name a few.

Among the 122-type materials, Mn-based $A$Mn$_2$$Pn_{2}$ ($A$ = Ca, Sr, Ba and $Pn$ = P, As, Sb, Bi) semiconducting compounds have received special attention. Irrespective of the type of pnictide ion present, the Ca- and Sr-based $A$Mn$_2$$Pn_2$ materials crystallize in CaAl$_2$Si$_2$-type trigonal structure with space group $P{\bar 3}m1$ while the Ba-based compounds crystallize in ThCr$_2$Si$_2$-type tetragonal structure, suggesting that alkaline earth metals are the controlling factor for establishment of the ground state crystal structure in these systems. Further, pnictide atoms control the unit cell dimensions, hence, the separation between magnetic Mn-ions \cite{Mewis-1978,Cordier-1976}. As a result, with two accessible composition-based tunable control parameters, these systems offer immense opportunities for the investigation of structure-interactions-property relationship. 

Specifically, small band gap semiconducting BaMn$_2$$Pn_{2}$ compounds with stacked-square-lattice of Mn ions undergo G-type AFM ordering with N\'eel temperature $T_{\rm N}$ considerably higher than room temperature (Table~\ref{Table:Parameters}). Interestingly, the most well-studied member of this family, BaMn$_2$As$_2$, undergoes insulator-to-metal transition by a small amount ($<2\%$) of hole-doping \cite{Pandey-2012} or by the application of moderate $\sim6$~GPa pressure \cite{Satya-2011}. Unlike the related itinerant AFM spin-density wave compound BaFe$_2$As$_2$, where hole-doping leads to the formation of SC ground state \cite{Paglione-2010,Johnston-2010,Stewart-2011}, the higher levels of hole-doping ($\gtrsim 40\%$) results in stimulating prototype half-metallic behavior in BaMn$_2$As$_2$ \cite{Pandey-2013a, Ueland-2015, Pandey-2015}. 

In contrast to the thoroughly investigated As-,Sb- and Bi-based counterparts \cite{Singh-2009,Sangeetha-2018,Saparov-2013,Calder-2014}, limited amount of experimental work has been reported on BaMn$_2$P$_2$. About two decades ago, Brock {\it et al.} performed magnetic susceptibility and neutron diffraction studies on a polycrystalline sample of BaMn$_2$P$_2$ and suggested that the material is a high-spin G-type antiferromagnet where Mn-spins carry a moment of 4.2(1)~$\mu_{\rm B}$ and $T_{\rm N} > 750$~K \cite{Brock-1994a,Brock-1994b}. After their work, there was no further attempt to investigate the properties of this material. In this paper we report the single crystal growth as well as structural, electronic, magnetic, and thermal properties of BaMn$_2$P$_2$ and attempt to establish structure-property relationship within insulating BaMn$_2$$Pn_{2}$ compounds. We show that the value of $T_{\rm N}$ in BaMn$_2$P$_2$ is highest among the 122-pnictide compounds. Additionally, our results also establish a strong correlation of the value of $T_{\rm N}$ on intralayer as well as interlayer distance between the Mn-ions.  

\section{Experimental Details}

Single crystals of BaMn$_2$P$_2$ were grown using solution growth technique in Sn flux using high purity starting elements Ba (99.99\%), Mn (99.98\%), P (99.999\%) and Sn (99.999\%) weighed in the molar ratio Ba:Mn:P:Sn = 1:2:2:45. The required amounts of the pure elements were placed in an alumina crucible and sealed inside a quartz tube under a partial Ar pressure of about 0.25 atm. The sealed tube was placed in a furnace and heated from room temperature to 400~$^\circ$C in 10~h and held at this temperature for 30~h. The assembly was further heated to 1180~$^\circ$C in 25~h and maintained at this temperature for 40~h. Further, it was slow-cooled to 700~$^\circ$C in 130~h. At this temperature, the Sn flux was decanted \cite{Canfield-2019} using a centrifuge and shiny plate-like crystals of BaMn$_2$P$_2$ of typical size of $2~\times2~\times 0.5~{\rm mm}^3$ were obtained. Room temperature powder x-ray diffraction (XRD) measurement was performed on a few small crushed crystals of BaMn$_2$P$_2$ using a Bruker D8 Advance Powder Diffractometer equipped with a Cu-K$_\alpha$ radiation source. Two-phase Rietveld refinement was performed using the FullPROF package \cite{Carvajal-1993}.

Temperature $T$ dependence of heat capacity $C_{\rm p}$ as well as four-probe basal $ab$-plane electrical resistivity $\rho_{ab}$ measurements were done using a Physical Properties Measurement System (PPMS) of Quantum Design Inc. (QDI), USA. Single component silver-filled epoxy was used for making electrical contacts of platinum wires on the single crystal for electrical measurements.
Temperature dependence of anisotropic magnetic susceptibilities $\chi_{ab,c}(T)$ were measured using two different options: for 2~K $\le T \le 350$~K using a SQUID-Magnetic Properties Measurement System (MPMS) of QDI and for 350~K$ \le T \le 850$~K using the high-temperature Vibrating Sample Magnetometer (VSM) Option of a PPMS. Isothermal magnetization $M$ versus applied field $H$ at five different temperatures were measured using MPMS. 

\begin{figure}
\includegraphics[width=3.1in]{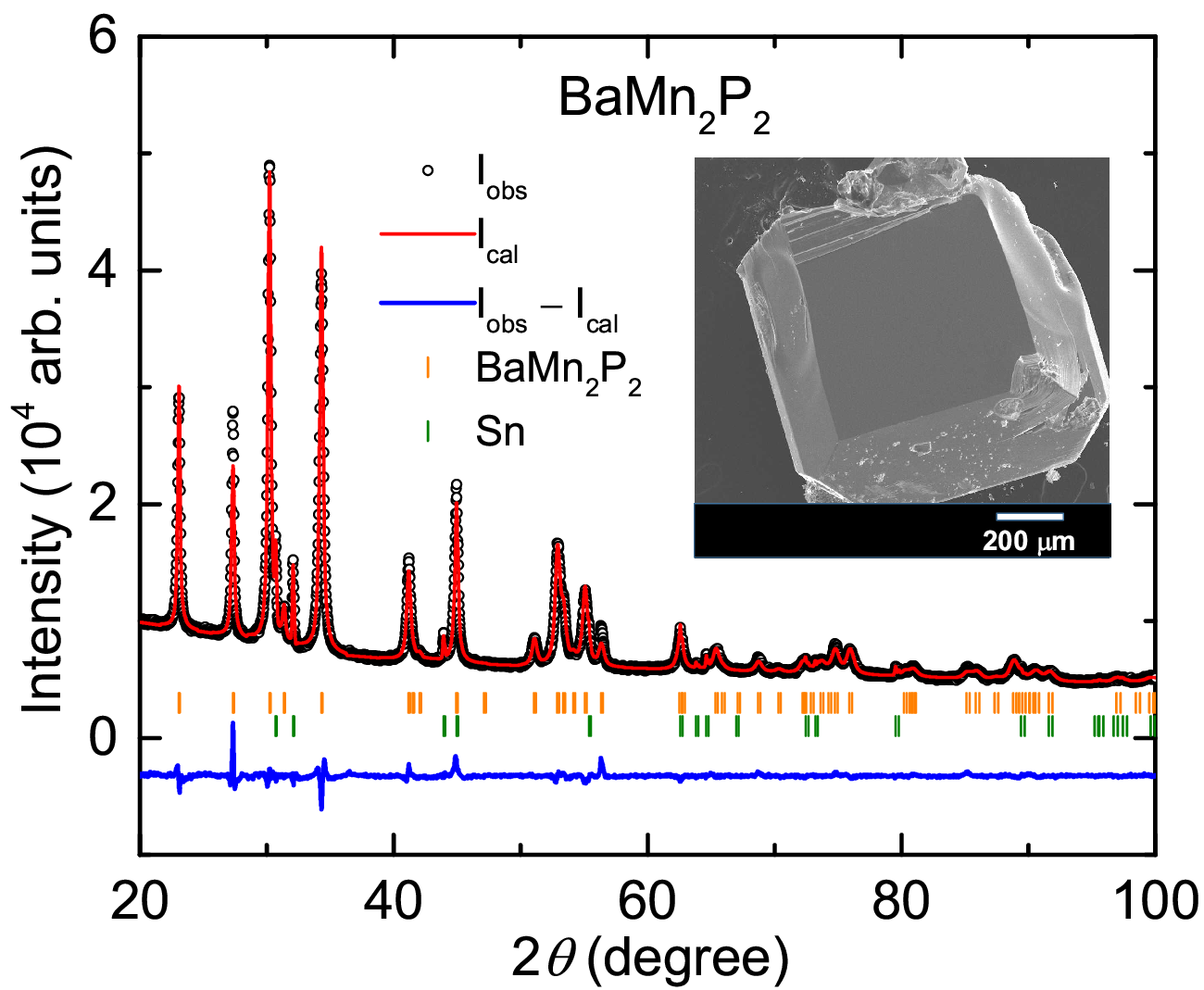}
\caption{Room temperature powder x-ray diffraction data collected on Sn-grown crushed crystals of BaMn$_2$P$_2$, two-phase Rietveld refinement profile, difference profile and Bragg positions of BaMn$_2$P$_2$ as well as adventitious Sn. Inset: scanning electron microscope image of a crystal of BaMn$_2$P$_2$.}
\label{fig:XRD}
\end{figure}

\begin{figure*}
\subfigure{\includegraphics[width=2.2in]{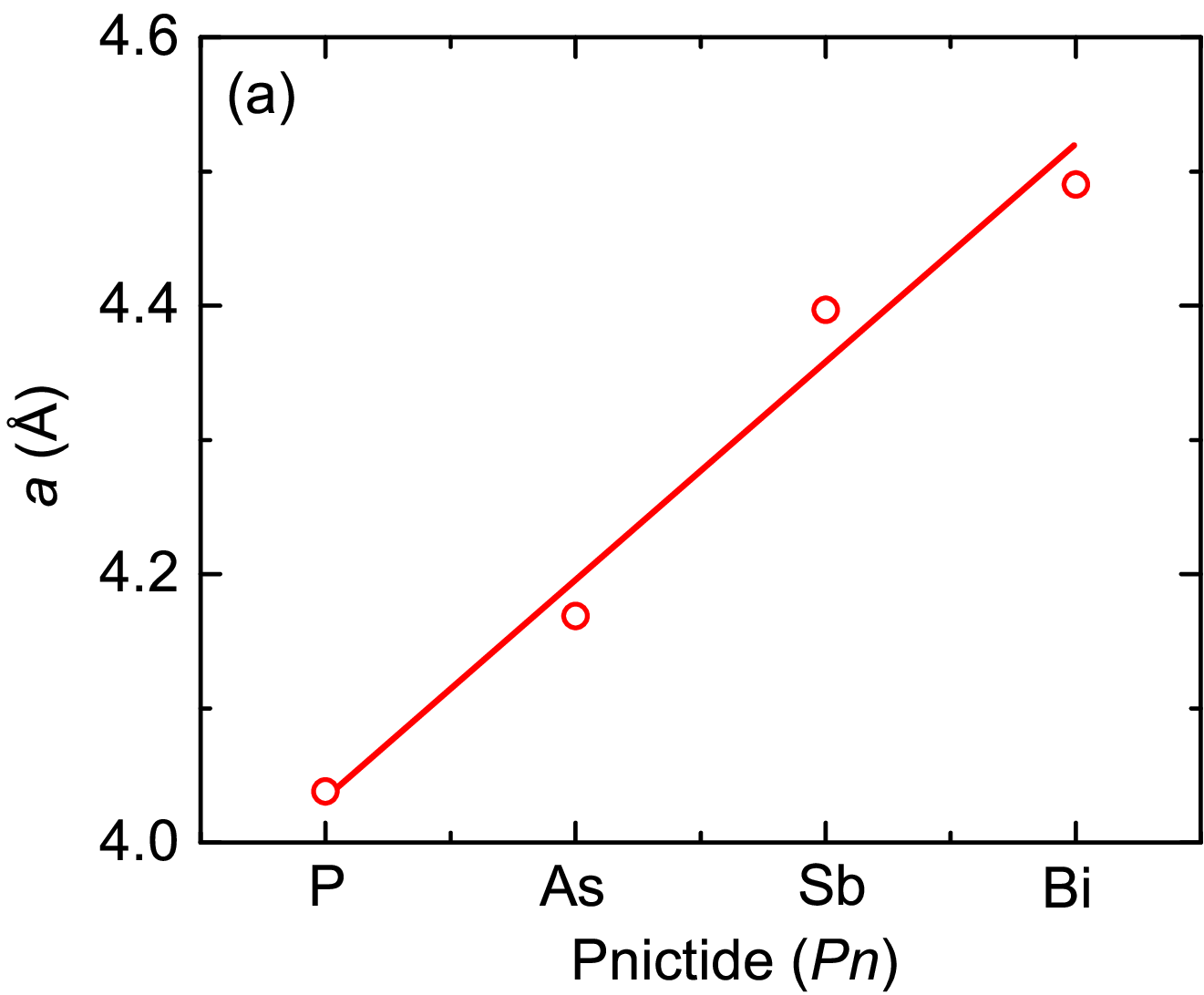}}
\subfigure{\includegraphics[width=2.2in]{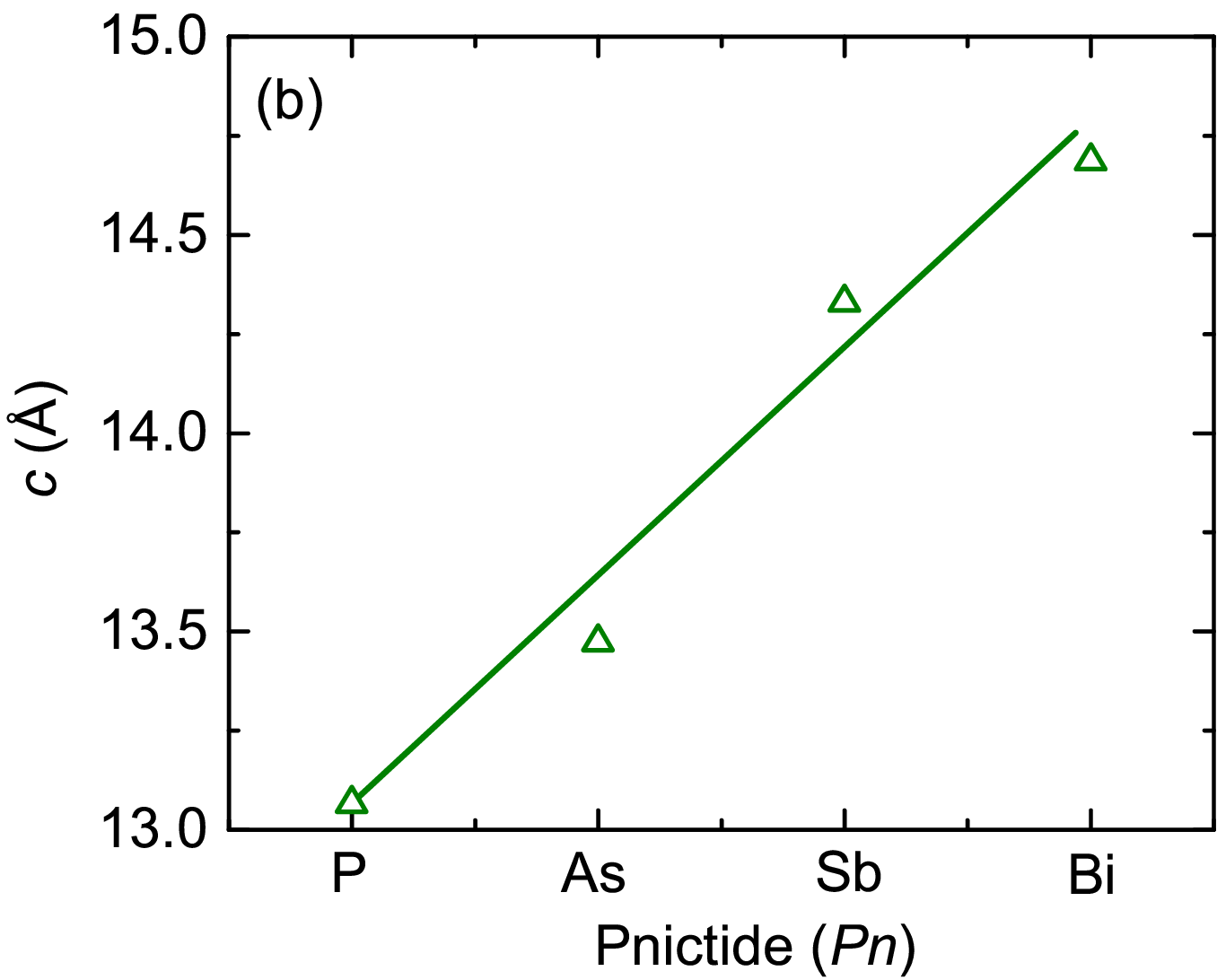}}
\subfigure{\includegraphics[width=2.2in]{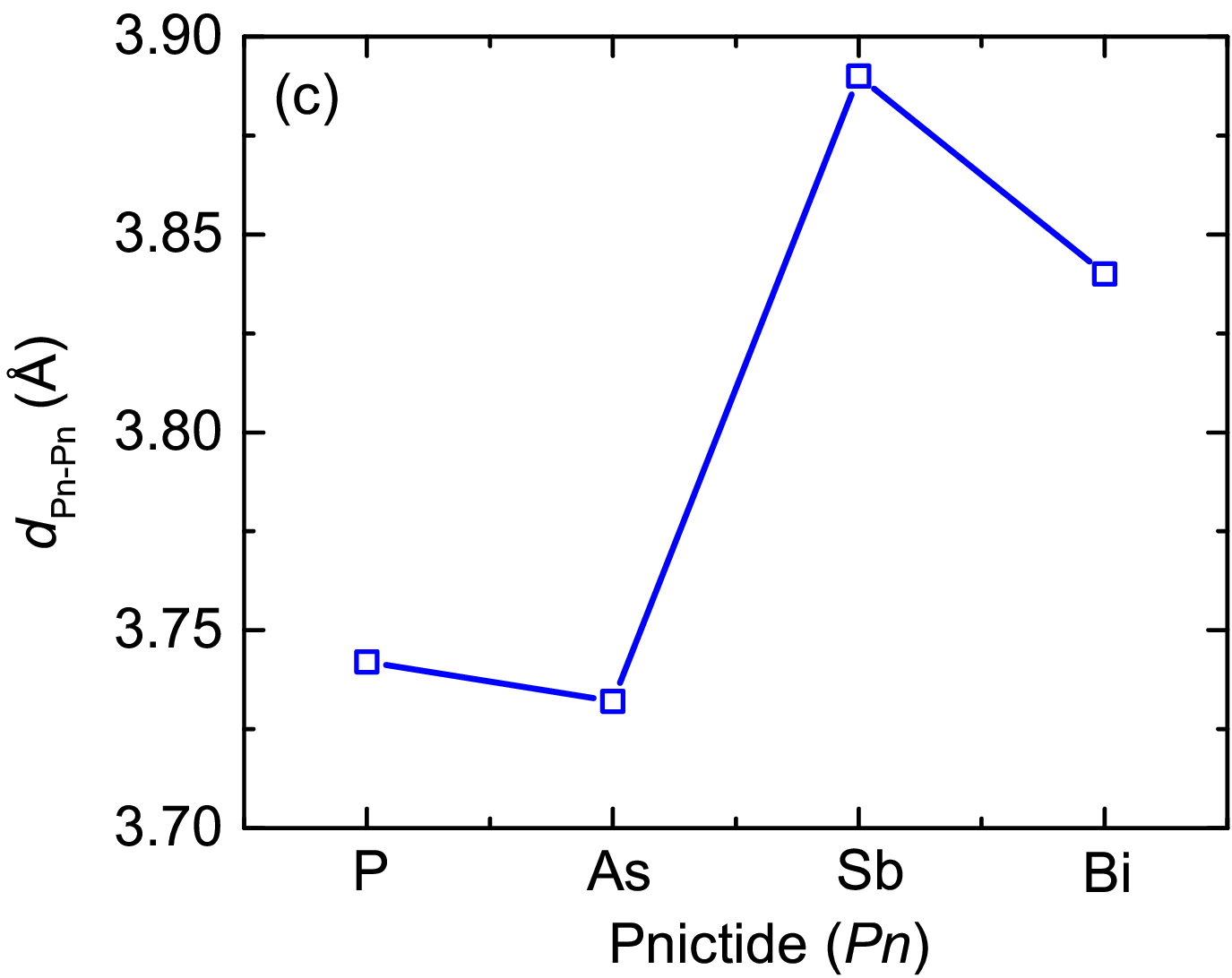}}
\caption{Variation of tetragonal lattice parameters (a) $a$ (b) $c$, and (c) the distance between adjacent pnictogen layers $d_{Pn-Pn}$ in BaMn$_2$$Pn_{2}$ ($Pn$ = P, As, Sb, Bi) with the type of pnictide ion present in the unit cell.}
\label{fig:LP}
\end{figure*}

\section{Results}

\subsection{A. Crystal structure}

The powder XRD data collected on a few crushed crystals of BaMn$_2$P$_2$ along with the two-phase Rietveld refinement fit, difference profile and Bragg positions are shown in the Fig.~\ref{fig:XRD}. The two-phase Rietveld refinement was performed to include the reflections originating from the small amount of adventitious Sn flux remaining on the surface of the small crystals that were crushed to powder. The fitted values of the tetragonal lattice parameters $a$, $c$ and the $c$-axis phosphorus position parameter $z_{\rm P}$ of BaMn$_2$P$_2$ are listed in Table~\ref{Table:Parameters} along with the corresponding parameters of BaMn$_2$As$_2$, BaMn$_2$Sb$_2$, and BaMn$_2$Bi$_2$. Figure~\ref{fig:LP}(a) and (b) show the variation of the lattice parameters $a$ and $c$, respectively, of the four BaMn$_2$$Pn_{2}$ compounds with the type of pnictogen atom present in the unit cell. The results highlight a nearly linear increase in $a$ and $c$ as we go down in the group from P to Bi. Due to the linear variation in the lattice parameters, the $c/a$ ratio of these materials stay nearly constant at a value of 3.25(2) (Table~\ref{Table:Parameters}). In contrast to the large variation of the $a$ and $c$ parameters, the distance between adjacent pnictogen layers $d_{Pn-Pn} = c(1-2z_{Pn})$ does not show any noteworthy dependence on the type of pnictogen present in the lattice and exhibits a nonmonotonic behavior with a mean value of 3.81(8)~\AA\ [Fig.~\ref{fig:LP}(c)]. It is also important to highlight here that the intralayer distances between nearest (NN)- and next-nearest-neighbor (NNN) Mn-spins are $a/\sqrt{2}$ and $a$, respectively, while the interlayer distance between the two adjacent Mn-spins along the $c$-axis is $c/2$. 

\begin{table*}[t]
\caption{Structural and physical properties parameters of BaMn$_2$$Pn_2$ ($Pn$ = P, As, Sb, Bi). All four compounds crystallize in ThCr$_2$Si$_2$-type tetragonal structure with space group $I4/mmm$ (SG: 221). The listed crystallographic parameters are the unit cell parameters $a$, $c$, $c/a$, $z_{Pn}$, and the distance between two nearest inter-layer pnictogen atoms $d_{Pn-Pn}$. The listed physical properties parameters are N\'eel temperature $T_{\rm N}$, Sommerfeld coefficient $\gamma$, coefficent $\beta$ and $\delta$ of lattice heat capacity, Debye temperature $\Theta_{\rm D}$, and band gap $\Delta$ obtained from the resistivity measurements.}
\label{Table:Parameters}
\begin{ruledtabular}
\begin{tabular}{l l l l l}
     Parameter & BaMn$_2$P$_2$ & BaMn$_2$As$_2$ & BaMn$_2$Sb$_2$  & BaMn$_2$Bi$_2$ \\
		           & (This work) & Ref.~\cite{Singh-2009} & Ref.~\cite{Sangeetha-2018} & Ref.~\cite{Saparov-2013} \\
	 \hline
			\underline{\bf Structure Parameters} &  &  &  & \\
			$a$(\AA) & 4.0381(1) & 4.1686(4) & 4.397(4) & 4.4902(3) \\
			$c$(\AA) & 13.0653(3) & 13.473(3) & 14.33(2) & 14.687(1)\\
			$c/a$ & 3.2355(2) & 3.232(1)& 3.259(8) & 3.2709(5)\\
			$z_{\rm Pn}$ & 0.3568(2) & 0.3615(3) & 0.3642(1) & 0.3692(3) \\
			$d_{Pn-Pn}$(\AA) & 3.742(5) & 3.732(9) & 3.89(1) & 3.84(1) \\
			
			& & & &\\
			\underline{\bf Property Parameters} & & & & \\
			$T_{\rm N}$ (K) & 795(15) & 618(3) & 450(10) & 387.2(4) \\
			$\gamma$ (mJ/mol~K$^2$) & 0.9(9) & 0.0(4)  & 0.02(4) & 0\\
			$\beta$ (mJ/mol~K$^4$)  & 0.23(3) & 0.65(3) & 0.680(3) & 0.434(3)\\
			$\delta$ (mJ/mol~K$^6$)  & 0.0027(1)& 0 & 0 & 0 \\
			$\Theta_{\rm D}$ (K) & 348(17) & 246(4) & 202(1)& 165 \\
			$\Delta$ (meV) & 24(2) & 27 \& 6.5  & 160 \& 18 & -- \\
\end{tabular}
\end{ruledtabular}
\end{table*}

\subsection{B. Electrical Transport}

\begin{figure}[t]
\includegraphics[width=3.1in]{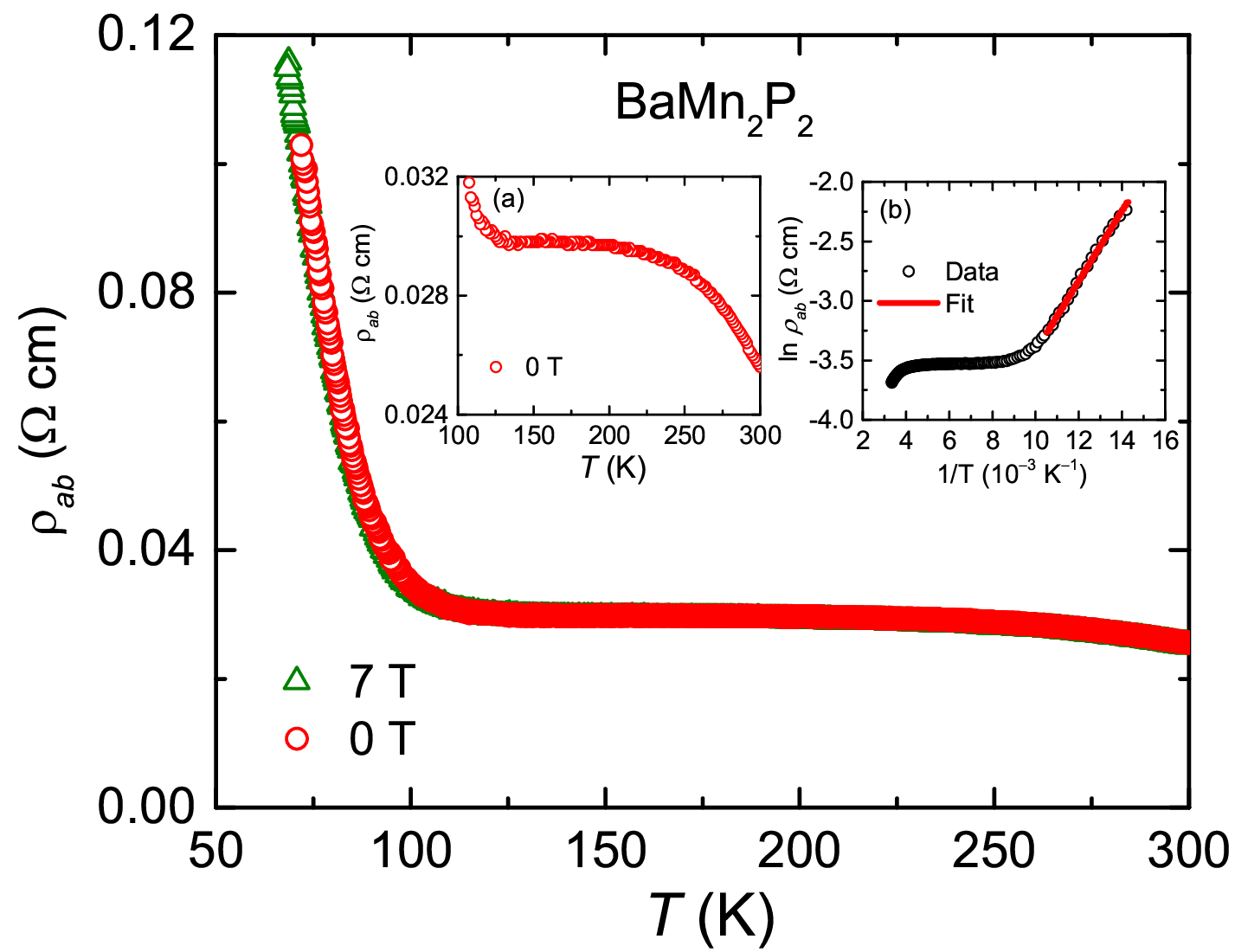}
\caption{Electrical resistivity $\rho_{ab}$ versus temperature $T$ of BaMn$_2$P$_2$ measured at two different applied magnetic fields $\mu_{0}H = 0$ and 7~T. Insets (a) $\rho_{ab}(T)$ for $100 \le T \le 300$~K and (b) ln$\rho_{ab}$ versus $T^{-1}$ along with a linear fit performed for 2~K$\le T\le 95$~K.\@ }
\label{fig:Rho}
\end{figure}

Figure~\ref{fig:Rho} shows the $ab$-plane electrical resistivity $\rho_{ab}$ of a single crystal of BaMn$_2$P$_2$ versus $T$ for two different applied fields $\mu_{0}H = 0$ and 7~T\@. The $\rho_{ab}(T)$ stays nearly constant and exhibits a small increase of about 10\% with decreasing $T$ from 300~K to $\sim 130$~K [Inset(a), Fig.~\ref{fig:Rho}]. However, there is a sharp rise in the $\rho_{ab}(T)$ values below 130~K where it shows a four fold increase and becomes non-measurable with the PPMS electronics below $\sim 70$~K. This observation clearly suggests that similar to the As-, Sb- and Bi-based analogs BaMn$_2$P$_2$ has an insulating ground state. Additionally, $\rho_{ab}(T)$ data collected at $\mu_{0}H = 0$ and 7~T overlap with each other in the entire $T$ range of measurement. This observation suggests that $\mu_{0}H \le 7$~T does not have any measurable effect on the interactions and scatterings that dominate the electrical transport within the $ab$-plane of this compound, leading to negligible magnetoresistance. The $\rho_{ab}(T)$ data below $\lesssim 100$~K show feature of activated transport [Inset(b), Fig.~\ref{fig:Rho}] and were fitted using
\begin{equation}
{\rm ln}\rho_{ab}(T) = A + \frac{\Delta}{k_{\rm B}T},
\label{eq:Res}
\end{equation}
where $A$ is a constant, $\Delta$ is activation energy of the band gap and $k_{\rm B}$ is Boltzmann constant. The fitted value of $\Delta = 24(2)$~meV suggesting it to be a small band gap material. This value is similar to the values reported on a polycrystalline sample of BaMn$_2$P$_2$ \cite{Brock-1994b} as well as single crystals of BaMn$_2$As$_2$ and BaMn$_2$Sb$_2$ (Table~\ref{Table:Parameters}). 

\subsection{C. Heat Capacity}

\begin{figure}
\includegraphics[width=3.1in]{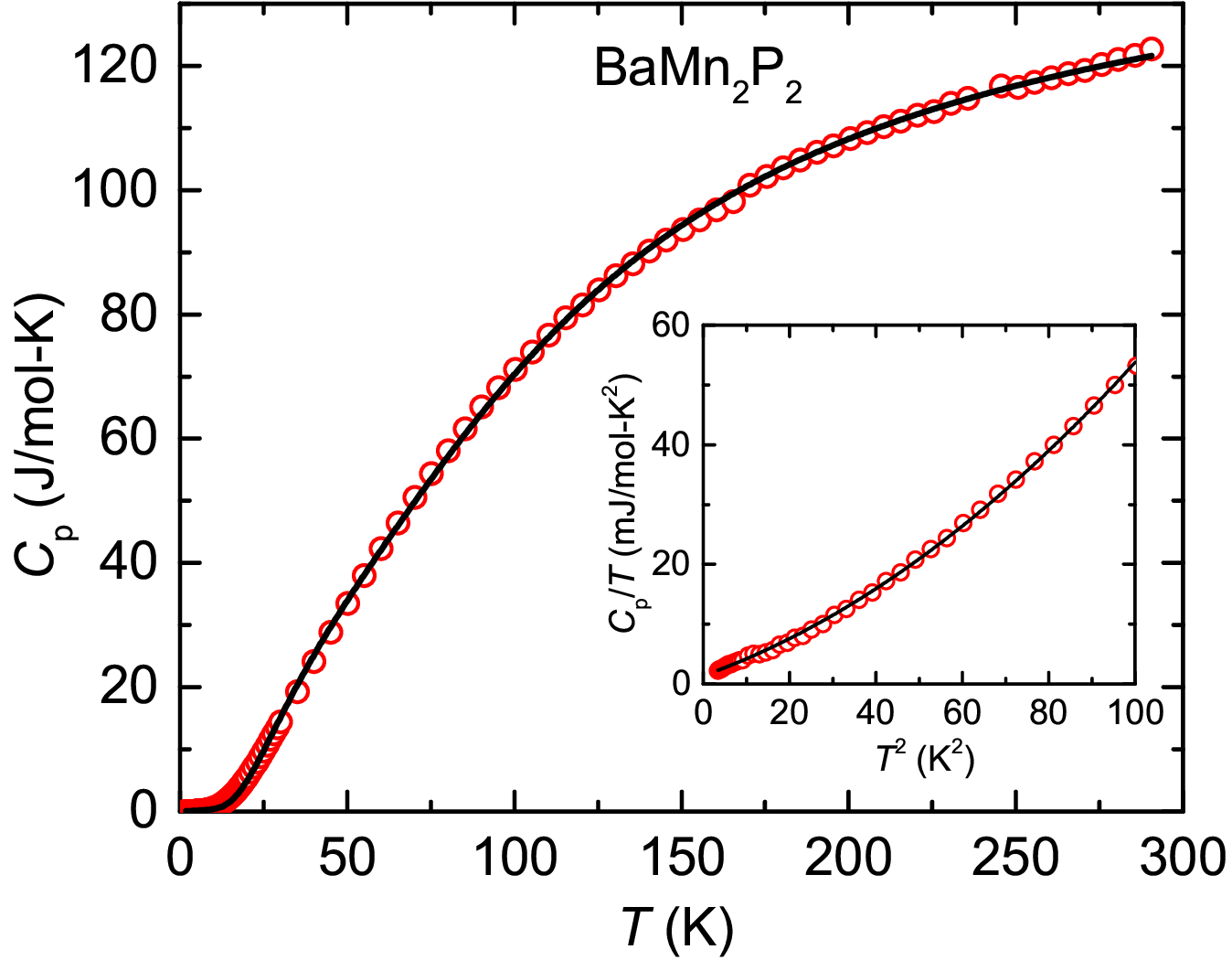}
\caption{Heat capacity $C_{\rm p}$ versus temperature $T$ along with the Debye-Einstein fit as described in the text. Inset: $C_{\rm p}/T$ versus $T^2$ data for $T\le 10$~K along with the fit as described in the text.}
\label{fig:HC}
\end{figure}

The heat capacity at constant pressure $C_{\rm p}$ versus $T$ data of BaMn$_2$P$_2$ do not show evidence of any phase transition below room temperature and attain a value of 122.7~J/mol\,K$^2$ at 291~K (Fig.~\ref{fig:HC}). This value is close to the Dulong-Petit high-$T$ limit of lattice heat capacity at constant volume given by $C_{V} = 3nR = 15R = 124.7$~J/mol\,K$^2$, where $n = 5$ is the number of atoms per unit cell and $R$ is the gas constant. Attempt to fit the $C_{\rm p}(T)$ data using Debye model of acoustic phonons \cite{Goetsch-2012,Gopal-1966} led to a poor agreement especially at intermediate temperatures. This observation is not surprising as Debye model works well both at low temperatures ($T \ll \Theta_{\rm D}$) where it predicts a $T^3$ dependence of $C_{\rm V}$ and at high temperatures ($T \gtrsim \Theta_{\rm D}$) where it leads to the Dulong-Petit limit. Here $\Theta_{\rm D}$ is Debye temperature of the material under consideration. However, the model often becomes inconsistent with the experimental results at intermediate temperatures. Often for materials with high $\Theta_{\rm D}$ one gets a significantly improved fit by incorporating a weighted Einstein single-frequency contribution to the Debye model as in; 
\begin{equation}
C_{\rm p}(T) = uC_{\rm{V\,Debye}}(T) +  (1-u)C_{\rm{V\,Einstein}}(T),
\label{eq:HC}
\end{equation}
where $C_{\rm{V\,Debye}}(T)$ and $C_{\rm{V\,Einstein}}(T)$ are the Debye and Einstein contributions, respectively, and $u$ is fraction of the Debye contribution to the lattice heat capacity \cite{Gopal-1966}. We got a good agreement to the $C_{\rm p}(T)$ data using Eq.~\ref{eq:HC} and the fitted value of $u$, $\Theta_{\rm D}$ and Einstein temperature $\Theta_{\rm E}$  were 0.324(6), 498(4) and 118(1)~K, respectively (Table~\ref{Table:Parameters}). As expected, the fitted value of the $\Theta_{\rm D}$ is higher than room temperature. 

The low temperature $C_{\rm p}(T)$ data were fitted using, 
\begin{equation}
C_{\rm p}/T =  \gamma + \beta T^2 + \delta T^4,
\label{eq:HC-lowT}
\end{equation}
where $\gamma$ is the Sommerfeld electronic heat capacity coefficient while $\beta$ and $\delta$ are the coefficients of low temperature lattice contributions. We obtained a good fit to the low-temperature $C_{\rm p}(T)$ data using Eq.~\ref{eq:HC-lowT} [Inset, Fig.~\ref{fig:HC}] and the fitted values of the parameters are; $\gamma = 0.9(9)$~mJ/mol K$^2$, $\beta = 0.23(3)$~mJ/mol K$^4$ and $\delta = 0.0027(1)$~mJ/mol K$^6$. As expected for a material with insulating ground state, the $\gamma$ is nearly zero indicating absence of density of states at Fermi level $E_{\rm F}$ at low temperatures. The low-$T$ $\Theta_{\rm D}$ calculated from the value of $\beta$ using $\Theta_{\rm D} = (12\pi^4Rn/5\beta)^{1/3} = 348(17)$~K is significantly higher than those reported for As-, Sb- and Bi-counterparts (Table~\ref{Table:Parameters}). 

\subsection{D. Magnetism}

\begin{figure}
\includegraphics[width=3.1in]{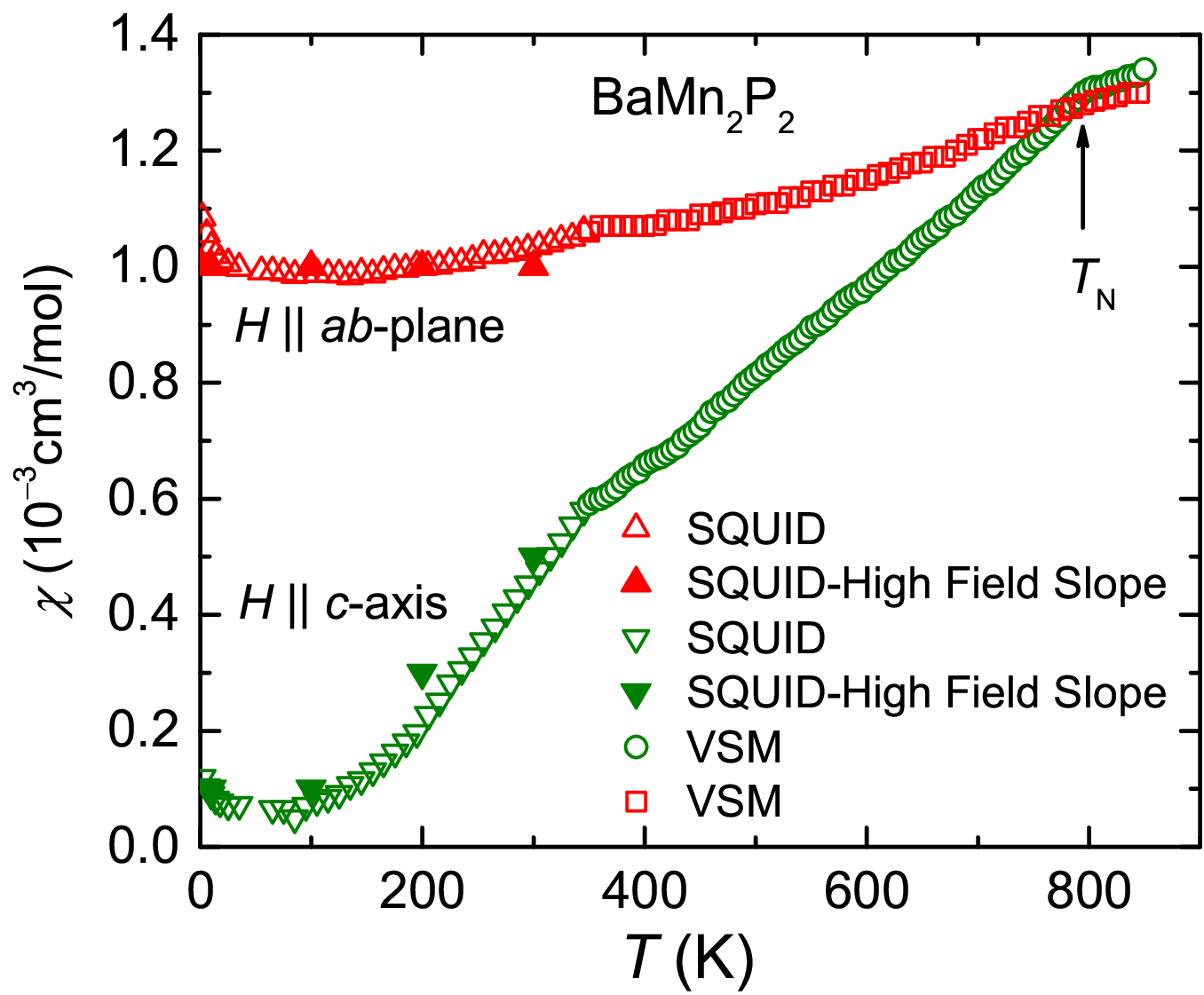}
\caption{Anisotropic magnetic susceptibilities $\chi_{ab,c} \equiv M/H$ versus temperature $T$ of BaMn$_2$P$_2$ measured using a SQUID-magnetometer at $\mu_{0}H = 3$~T for $T \le 350$~K, determined from high-field slopes of the isothermal magnetization measurements and measured using a vibrating sample magnetometer for $T \ge 350$~K\@. }
\label{fig:MT}
\end{figure}

\begin{figure*}
\subfigure{\includegraphics[width=2.2in]{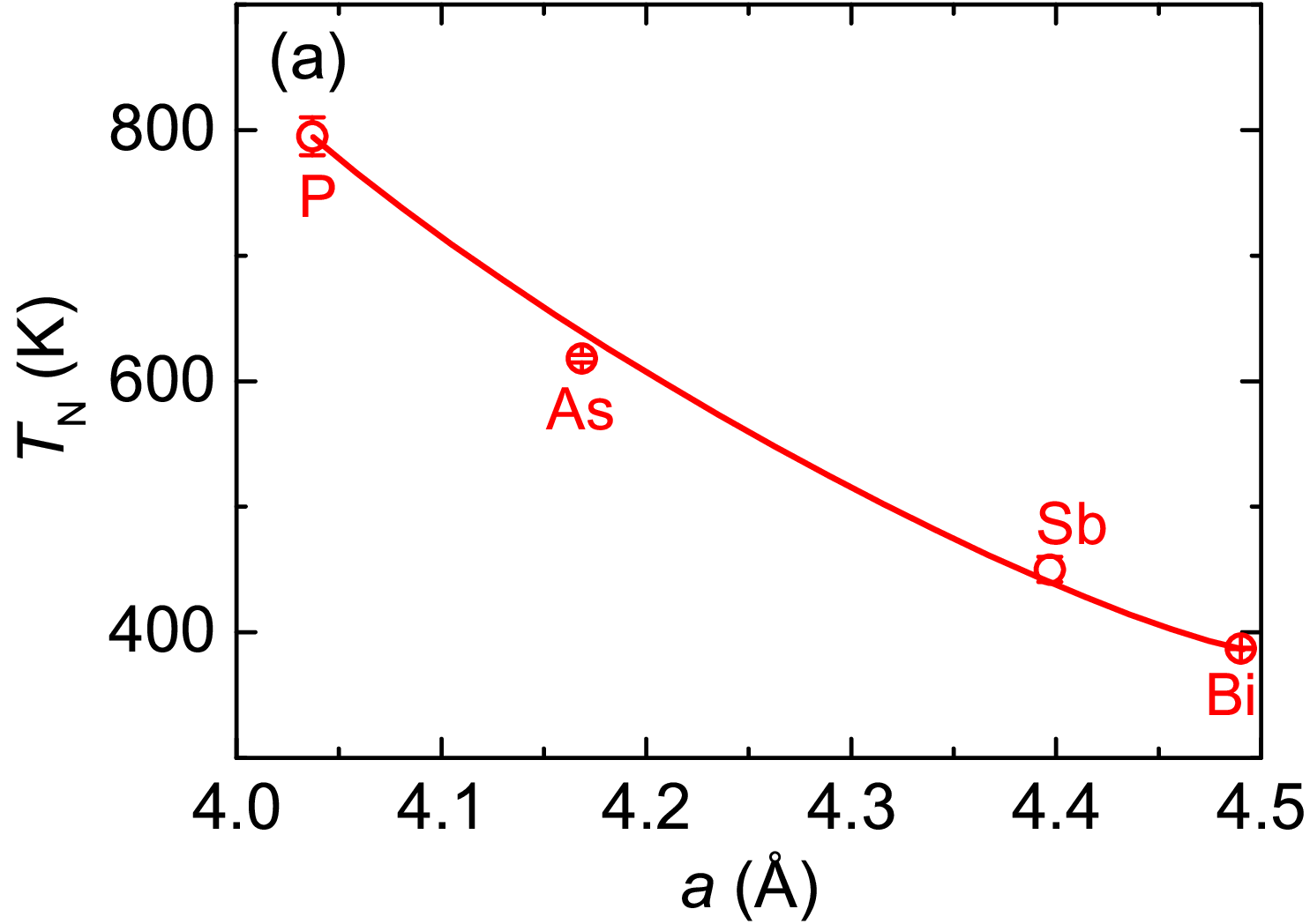}}
\subfigure{\includegraphics[width=2.2in]{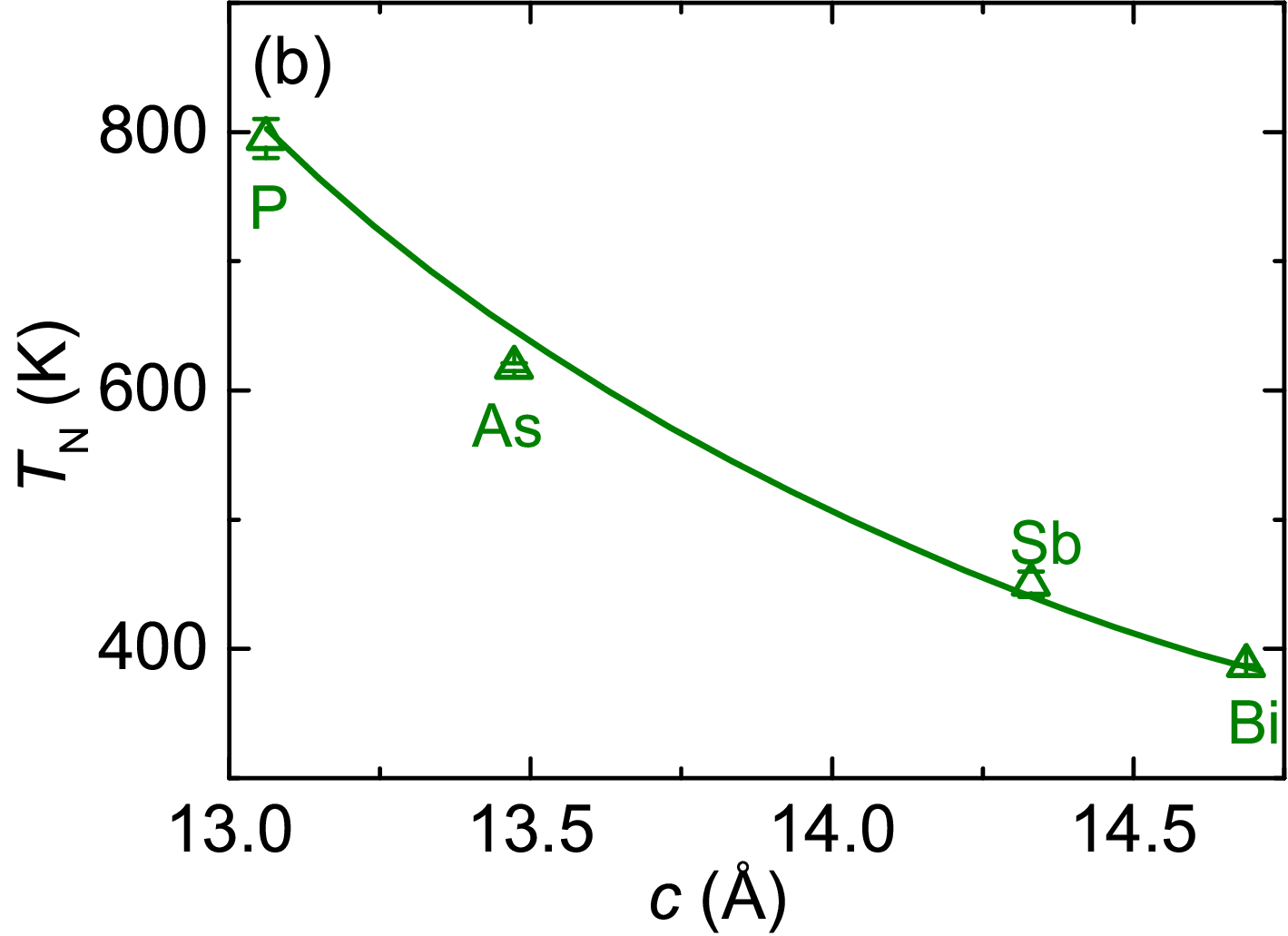}}
\subfigure{\includegraphics[width=2.2in]{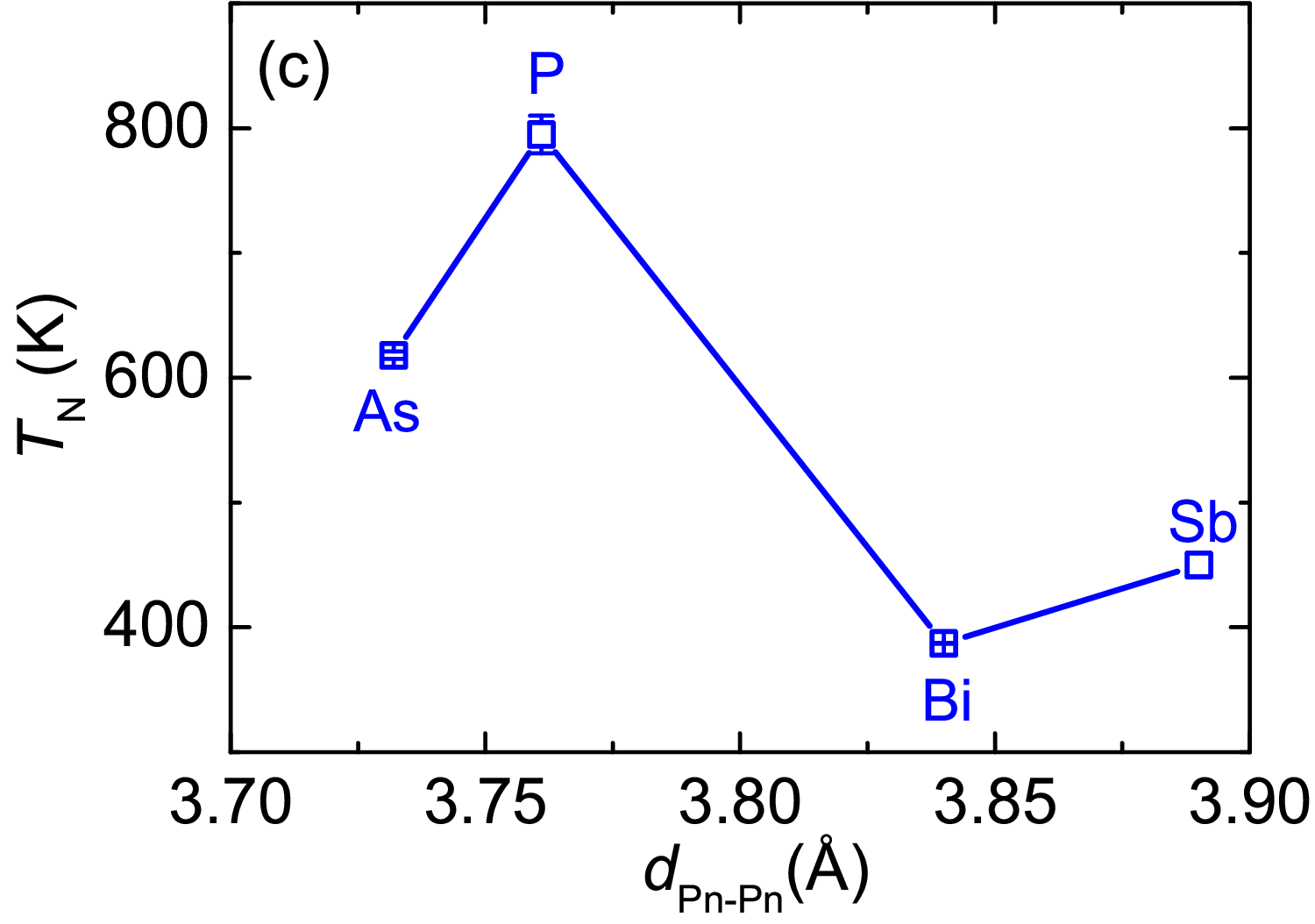}}
\caption{Variation of antiferromagnetic transition temperature $T_{\rm N}$ with tetragonal unit cell parameters (a) $a$ (b) $c$ and (c) distance between adjacent pnictogen layers $d_{Pn-Pn}$ in BaMn$_2$$Pn_2$ ($Pn$ = P, As, Sb, Bi).}
\label{fig:LP-TN}
\end{figure*}

Temperature dependence of the magnetic susceptibilities $\chi_{ab}$ ($H \parallel ab$-plane) and $\chi_{c}$ ($H \parallel c$-axis) of a crystal of BaMn$_2$P$_2$ measured at $\mu_{0}H = 3$~T are shown in Fig.~\ref{fig:MT}. The $\chi(T)$ data taken using a SQUID magnetometer for $T\le 400$~K and those collected using a PPMS-VSM for $T\ge 350$~K connect nearly smoothly, leaving a minor kink at 350~K that likely arises due to slight variation in the orientations of the crystal relative to $H$ when mounted on two different set-ups. The SQUID data are shown up to 350~K for clarity of the figure. The observed $T$-dependence of the anisotropic $\chi_{ab, c}$ data suggests that similar to the As-, Sb-, and Bi-based analog compounds, BaMn$_2$P$_2$ has a collinear AFM ordering where the localized spins are oriented along the $c$-axis of the tetragonal unit cell. The $T_{\rm N} = 795(15)$~K deduced from the $\chi_{ab, c}(T)$ data is considerably higher than that observed in the other three materials, suggesting the presence of comparatively stronger magnetic interactions in BaMn$_2$P$_2$. Further, the $\chi_{ab, c}(T)$ data show a monotonic increase above $T_{\rm N}$ suggesting the persistence of dynamic short-range quasi-two dimensional AFM correlations above the ordering temperature, a behavior which was earlier observed in the other three compounds discussed above. The results show that $T_{\rm N}$ of BaMn$_2$$Pn_{2}$ ($Pn$ = P, As, Sb, Bi) exhibits a monotonic decrease with increase in the tetragonal unit cell parameters $a$ and $c$ [Figs.~\ref{fig:LP-TN}(a) and (b)]. This observation infers the progressive weakening of the magnetic interactions operating between the localized Mn-spins with increase of the separation between them.  While moving from BaMn$_2$P$_2$ to BaMn$_2$Bi$_2$ we observe nearly 50\% reduction in $T_{\rm N}$ by approximately 12\% increase in the $a$ and $c$ (Table~\ref{Table:Parameters}). Our observations also clearly demonstrate that the $T_{\rm N}$ values do not have any direct correlation with the $d_{Pn-Pn}$ in these materials [Fig.~\ref{fig:LP-TN}(c)]. 

\begin{figure*}
\subfigure{\includegraphics[width=3.1in]{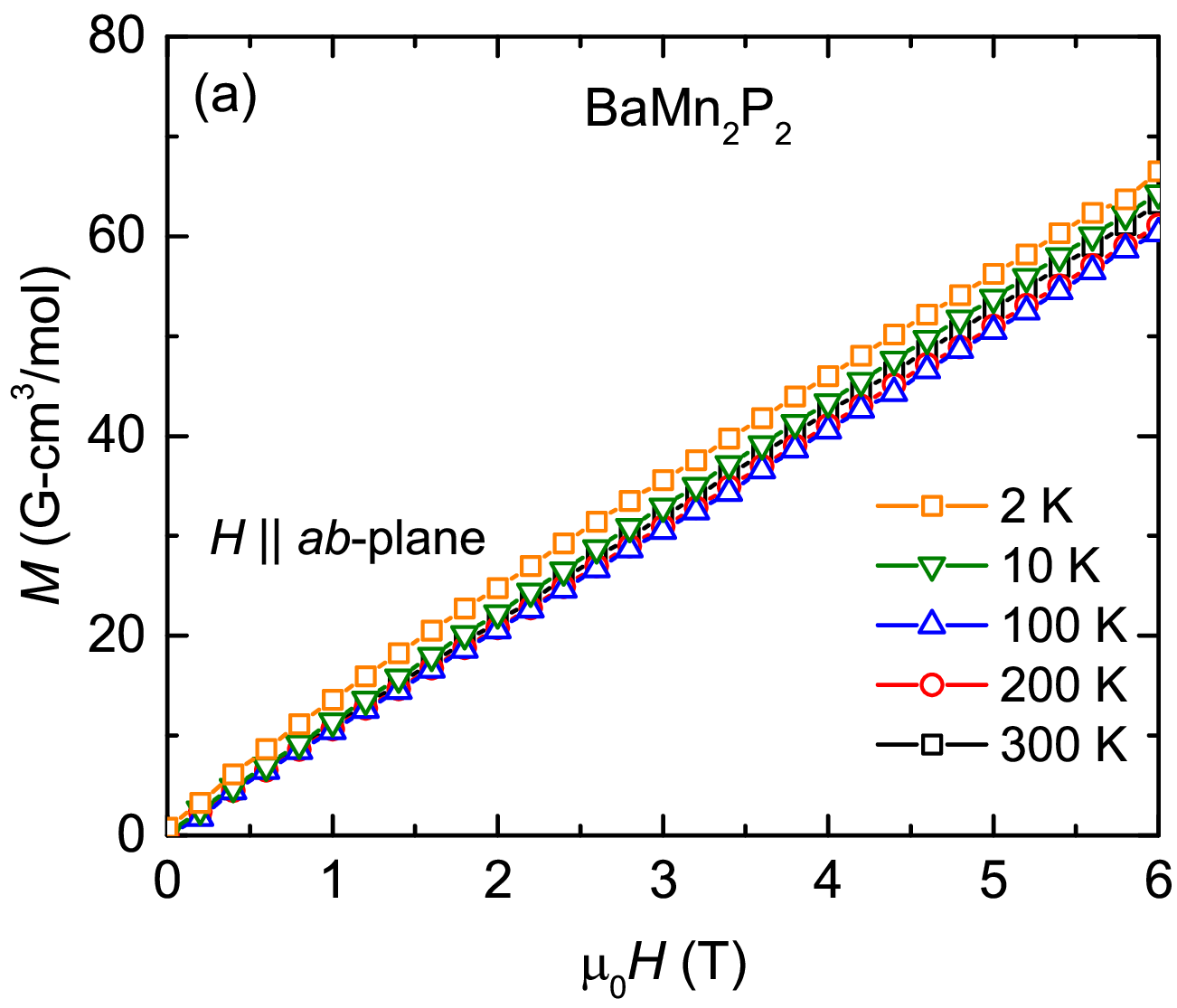}}
\subfigure{\includegraphics[width=3.1in]{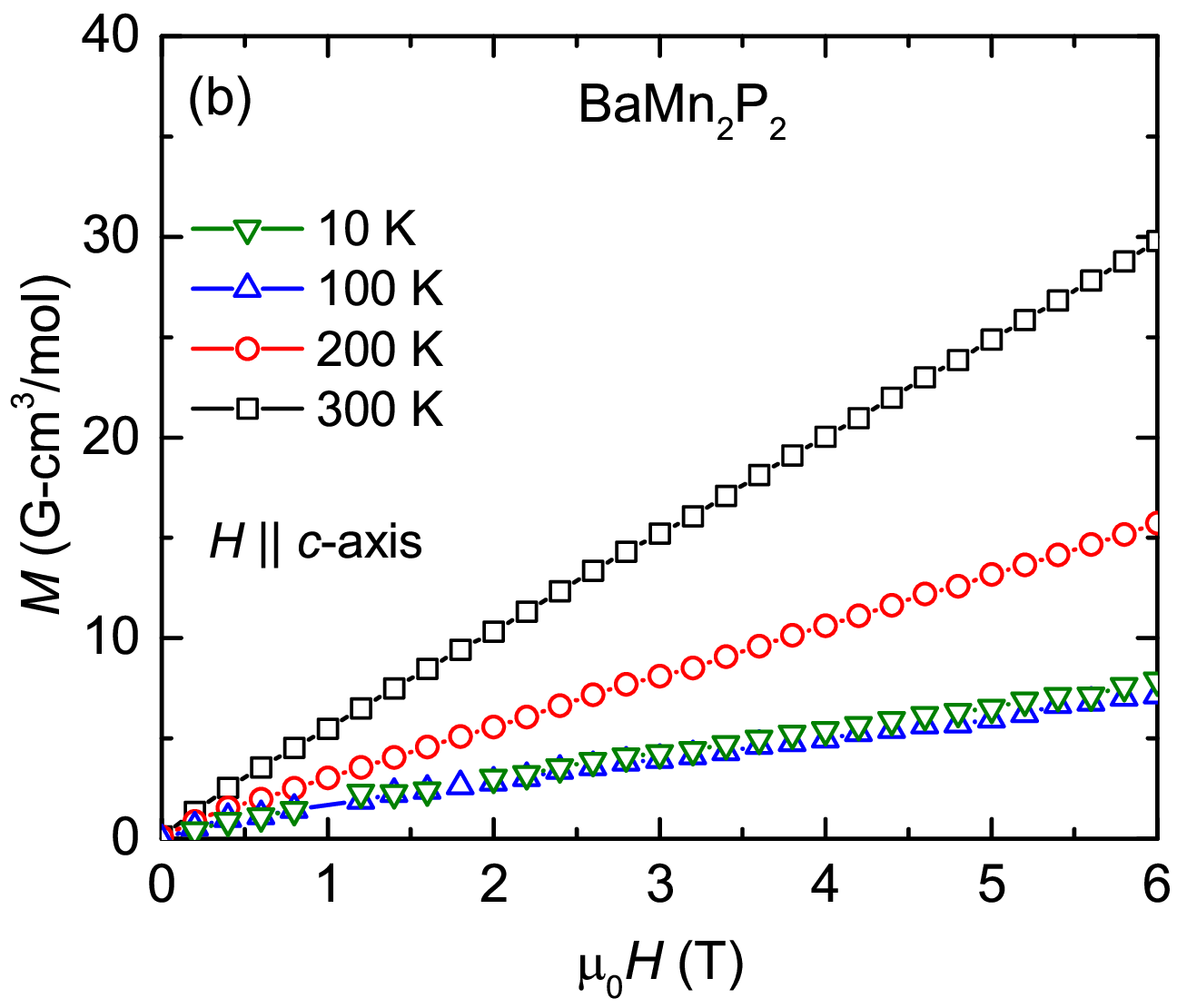}}
\caption{Isothermal magnetization $M$ versus applied magnetic field $H$ of BaMn$_2$P$_2$ measured for two different crystal orientations; (a) $H \parallel ab$-plane and (b) $H \parallel c$-axis.}
\label{fig:MH}
\end{figure*}

Isothermal magnetization $M$ of a single crystal of BaMn$_2$P$_2$ versus applied magnetic field $H$ for two orientations of the applied field $H \parallel ab$-plane and $H \parallel c$-axis are shown in Figs.~\ref{fig:MH}(a) and \ref{fig:MH}(b), respectively. As expected for a local-moment AFM system below its ordering temperature, the $M$ is proportional to $H$ for both orientations of the applied field for $T\le 300$~K\@. The slight non-linearity in the $M_{ab}(H)$ data at 2~K indicates the presence a very small amount of saturable paramagnetic or FM impurities in the crystal. The same was inferred by the small upturn observed in the $\chi_{ab}(T)$ below $\sim 10$~K (Fig. \ref{fig:MT}). Evidently the effect of these impurities is almost negligible in the $T$-range of our measurements. However, we still attempted to get the intrinsic susceptibilities at low temperatures from the high-field $\mu_{0}H\ge 4$~T slopes of the $M(H)$ data as shown by the solid symbols in Fig.~\ref{fig:MT}. This observation confirms that the minor upturn in $\chi_{ab}(T)$ below $\sim 10$~K is indeed due to the presence of trace saturable impurities.    

\section{Discussion}

Electrical transport measurements infer an insulating ground state in BaMn$_2$P$_{2}$ with activation energy $\Delta = 24(2)$~meV. Similar observations were earlier made on BaMn$_2$As$_{2}$ and BaMn$_2$Sb$_{2}$ (Table~\ref{Table:Parameters}). The $\rho_{ab}(T)$ of BaMn$_2$P$_{2}$ decreases exponentially with the increase of $T$ until $\sim 130$~K and then remains nearly $T$-independent up to $\sim 200$~K beyond which it decreases slightly with increase of $T$. This crossover indicates that similar to the As- and Sb-based analogs there is a possibility of the existence of more than one activation energies likely associated with different zones of the electronic bands dominating the electrical transport in different $T$ regions. The $C_{\rm p}(T)$ data lead to $\gamma = 0.9(9)$~mJ/mol~K$^2$ corroborating the outcome of the electrical transport measurement and infer an insulating ground state in BaMn$_2$P$_{2}$. The $\Theta_{\rm D}$ of BaMn$_2$$Pn_{2}$ compounds decreases rapidly with the increase in atomic number of pnictide atom present in the lattice (Table~\ref{Table:Parameters}) inferring that the higher energy modes get enhanced with the inclusion of lighter elements in the lattice.  

The insulating ground state of BaMn$_2$P$_{2}$ confirms that, similar to other BaMn$_2$$Pn_{2}$ compounds, the underlying magnetism in this material is constituted by the localized spins residing on Mn-ions. The anisotropic $\chi_{ab,c}(T)$ of BaMn$_2$P$_{2}$ clearly show the existence of a collinear AFM with the magnetic easy-axis parallel to the tetragonal $c$-axis and $T_{\rm N} = 795(15)$~K\@. The observed $T_{\rm N}$ in BaMn$_2$P$_{2}$ is highest among all the ThCr$_2$Si$_2$- and CaAl$_2$Si$_2$-type 122-pnictide compounds reported so far suggesting that the strength of magnetic exchange interactions is strongest in this material. The magnetic interactions within BaMn$_2$As$_{2}$ and BaMn$_2$Sb$_{2}$ have been successfully explained using $J_{1}$-$J_{2}$-$J_{c}$ Heisenberg model of stacked-square lattice, where $J_{1}$ and $J_{2}$ are the NN and NNN intralayer exchange interactions, respectively, and $J_{c}$ in the interlayer interaction acting along the $c$-axis \cite{Johnston-2011}. The similarity of the $\chi_{ab,c}(T)$ data of BaMn$_2$P$_{2}$ with those of the As-, Sb- and Bi-analogs \cite{Singh-2009,Sangeetha-2018,Saparov-2013} suggest that magnetic interactions within this material can also be described using the same model. Interestingly, we have observed a monotonic decrease of $T_{\rm N}$ with the increase in $a$ and $c$ parameters. When moving from BaMn$_2$P$_{2}$ to BaMn$_2$Bi$_{2}$, about $12\%$ increase in $a$ and $c$ leads to about $\sim 50$\% decrease in $T_{\rm N}$ suggesting a strong dependence of $J_{1}$, $J_{2}$ and $J_{c}$ on the intralayer as well as interlayer distances between the localized Mn-spins. Further, the $\chi_{ab,c}(T)$ of BaMn$_2$P$_{2}$ show a monotonic increase for $T>T_{\rm N}$. This behavior is similar to what was previously observed in As-, Sb- and Bi-analogs \cite{Sangeetha-2018,Saparov-2013,Johnston-2011} and suggests that dynamic short-range quasi two-dimensional AFM correlations persist above $T_{\rm N}$ in this material as well. 

It is interesting to compare the Mn-based insulating BaMn$_2$$Pn_{2}$ compounds with the Co-based metallic $A$Co$_2$As$_2$ ($A =$ Ca, Sr, Ba) which also crystallize in ThCr$_2$Si$_2$-type structure. $A$Co$_2$As$_2$ compounds exhibit properties that  delicately depend upon the interlayer $d_{\rm As-As}$, which  regulates the oxidation state of Co-ions in these materials by controlling the extent of the interlayer As-As bonds. As a result, it indirectly controls the magnetic ground state \cite{Pandey-2013b}. For example, BaCo$_2$As$_2$ with $d_{\rm As-As} = 3.78$~\AA\ shows a nonmagnetic ground state \cite{Anand-2014a}. On the other hand, CaCo$_{1.86}$As$_2$ with $d_{\rm As-As} = 2.73$~\AA\ exhibits an A-type collinear AFM ordering below 52~K \cite{Anand-2014d}. Interestingly, SrCo$_2$As$_2$ with $d_{\rm As-As} = 3.33$~\AA\ intermediate to those of its Ca and Ba analogs shows no evidence of long-range magnetic ordering, but exhibits stripe-type AFM \cite{Jayasekara-2013} as well as FM \cite{ Li-2019} spin fluctuations, where the former is similar to the fluctuations observed in the FeAs-based superconductors. On the contrary, $d_{Pn-Pn}$ does not show any significant variation within the BaMn$_2$$Pn_{2}$ compounds and because of the localized nature of $d$-bands it apparently does not have any direct effect on the oxidation state of the Mn-ions as well as on the magnetic ground state of these materials. $A$Co$_2$As$_2$ and BaMn$_2$$Pn_{2}$ compounds together present a text-book type example of how the structural parameters and electronic band structure can indirectly govern the magnetism within crystalline systems. It is also noteworthy that the ground state crystal structure of $A$Mn$_2$$Pn_{2}$ compounds only depends upon the type of alkaline-earth metal present in the lattice while the unit cell dimensions are controlled by the pnictogen atoms constituting the lattice. This feature could further be used to investigate the structural crossover regions to understand structure-property relationship within this system and to explore instability-driven phase-transitions and ground states. 

As BaMn$_2$P$_{2}$ has the highest $T_{\rm N}$ within this series, it would be very interesting to perform hole-doping studies on this material. A similar investigation in the related compound BaMn$_2$As$_{2}$ with $T_{\rm N} = 618$~K has led to the discovery of a prototype itinerant half-metallic ferromagnetism with $T_{\rm c} \approx 100$~K. It would be worth to investigate if the stronger exchange interactions resulting in large $T_{\rm N}$ in BaMn$_2$P$_{2}$ can assist in developing a half-metallic FM state with higher $T_{\rm c}$. Such investigations would be important as the high-$T_{\rm c}$ half-metals are desirable candidates for spin-polarized device applications.  
    
\section{Conclusion}

High-quality single crystals of BaMn$_2$P$_{2}$ were grown using Sn-flux. The $ab$-plane electrical resistivity as well as the heat capacity measurements suggest a small band gap insulating ground state of the material. The anisotropic magnetic susceptibility $\chi_{ab,c}(T)$ measurements show that similar to BaMn$_2$$Pn_{2}$ ($Pn$ = As, Sb, Bi) compounds, BaMn$_2$P$_{2}$ has a collinear N\'eel-type AFM ground state with transition temperature $T_{\rm N} = 795(15)$~K, which is highest among the 122-pnictide compounds investigated so far. The AFM transition temperatures of BaMn$_2$$Pn_{2}$ compounds show a monotonic decrease with the increase of tetragonal $a$ and $c$ parameters, where about $12\%$ increase in $a$ and $c$ leads to $\sim 50$\% decrease in $T_{\rm N}$. These observations suggest a strong dependence of the strength of the magnetic exchange interactions on the separation between the localized spins residing on the Mn-ions. The $\chi_{ab,c}(T)$ of BaMn$_2$P$_{2}$ show a monotonic increase for $T>T_{\rm N}$ suggesting the presence of sizable short-range dynamic quasi two-dimensional AFM correlations above the $T_{\rm N}$. Because of the high value of $T_{\rm N}$ of BaMn$_2$P$_{2}$, hole-doping investigations are highly sought for. Similar studies in related BaMn$_2$As$_{2}$ with $T_{\rm N} = 618$~K have led to the discovery of prototypical itinerant half-metallic behavior of the doped holes with $T_{\rm c} \approx 100$~K. Such investigations become more significant as half-metals with high-$T_{\rm c}$ are emerging as desirable candidates for spin-polarized transport-based applications.

{\setlength{\parindent}{1in}{\bf Acknowledgments}}\\

AP acknowledges the financial support from FRC and URC of the University of the Witwatersrand and National Research Foundation of South Africa. The work at Texas A\&M University was supported by the National Science Foundation through Grant No. NSF-DMR-1807451.

\end{document}